\documentclass[11pt]{article}
\usepackage[latin1]{inputenc}
\usepackage[english]{babel}
\usepackage[namelimits]{amsmath}
\usepackage{amssymb}
\usepackage{amsmath}
\usepackage{amsthm}
\usepackage{authblk}

\begin{document}
\title{{\bf  The 3D perturbed Schr\"{o}dinger Hamiltonian in a Friedmann flat spacetime testing the primordial universe in a non commutative spacetime}}
\author[1,2]{S. Fassari\thanks{sifassari@gmail.com}}
\author[1,2]{F. Rinaldi\thanks{f.rinaldi@unimarconi.it}}
\author[1,3,4]{S. Viaggiu\thanks{s.viaggiu@unimarconi.it and viaggiu@axp.mat.uniroma2.it}}
\affil[1]{Dipartimento di Fisica Nucleare, Subnucleare e delle Radiazioni, Universit\'a degli Studi Guglielmo Marconi, Via Plinio 44, I-00193 Rome, Italy}
\affil[2]{CERFIM, PO Box 1132, Via F. Rusca 1, CH-6601 Locarno, Switzerland}
\affil[3]{Universit\`a di Roma ``Tor Vergata'', Via della Ricerca Scientifica, 1, I-00133 Roma, Italy.}
\affil[4]{INFN, Sezione di Napoli, Complesso Universitario di Monte S. Angelo, Via Cintia Edificio 6, 80126 Napoli, Italy.}
\date{\today}
\maketitle
\begin{abstract}
In this paper we adapt the mathematical machinery presented in \cite{P1} to get, by means of the Laplace-Beltrami operator, the discrete spectrum of the Hamiltonian of the Schr\"{o}dinger operator perturbed by an actractive 3D delta interaction in a Friedmann flat universe. In particular,
as a consequence of the treatment in \cite{P1} suitable for a Minkowski spacetime, 
the discrete spectrum of the ground state and the first exited state in the above mentioned cosmic framework can be regained. Thus, the coupling constant $\lambda$ must be choosen as a function of the cosmic comooving time $t$ as ${\lambda}/a^{2}(t)$, with ${\lambda}$ be the one of the static Hamiltonian studied in \cite{P1}. In this way we can introduce a time dependent delta interaction which is relevant in a primordial universe, where $a(t)\rightarrow 0$ and becomes negligible at late times, with $a(t)>>1$. We investigate, with the so obtained model, quantum effects provided by point interactions in a strong gravitational regime near the big bang.
In particular, as a physically interesting application, we present a method to depict, in a semi-classical approximation, a test particle in a (non commutative) quantum spacetime where, thanks to Planckian effects,
the initial classical singularity disappears and as a consequnce a ground state with negative energy emerges. Conversely, in a scenario where the scale factor $a(t)$ follows the classical trajectory, this ground state is instable and thus physically cannot be carried out. 
\end{abstract}

\section{Introduction}
A longstanding theoretical problem is to unify the general relativity and quantum mechanics (see \cite{1} and reference therein).
Unfortunately, a complete quantum gravity theory is still lacking. In particular, difficulties are due to the covariance of general relativity  and to the almost impossible task, thanks to the equivalence principle, to define localized energies.\\
The main issue is due to the lack of an experimental signature as a guidance for theoretical physicists.
It is thus important and very useful to study possible detectable effects induced by gravity on given quantum system.
The most used approach is based on the Wheeler-DeWitt \cite{2}
equation, however non trivial yet solved problems concern the probabilistic intepretation of the wave function of the entire universe.
A more simple approach is to consider effects of the spacetime curvature on a test particle. This line of research has begun with the paper in \cite{I1}, where the results present in \cite{P1,I1,I2,I3,I4, I5} have been applied to a spacetime with a topological defect.
In \cite{P1,I1,I2,I3,I4,I5}, the mathematical rigorous treatment regarding the renormalization technique has been used in order to obtain the energy spectrum of free Hamiltonians with attractive delta interactions. 
Delta interactions provide a very useful modeling with wide and increasing applications in many areas of modern mathematical-physics framework (see for example \cite{I6,I7} and reference therein). In particular, in \cite{I7}, it has been shown that weakly interacting bosons of any arbitrary potential can be approximated using delta functions (see also \cite{I6}).
To this purpose, we may think to a physical situation near the classical big bang singularity where spacetime quantum fluctuations together with interactions among particles \cite{I6,I7} can be studied with a simplified 
model with our test particle interacting with a delta potential.
In this framework, as will be shown in this paper, Planckian fluctuations due to a non commutative spacetime \cite{I8,I9,I10,I11,I12,I13} give a negative contributon to the energy content of the universe. The above negative contribution can, in the present model, be obtained in terms of an attractive delta interactions, experienced by our test particle.
\\
In section two we present the attractive delta interactions in a Friedmann flat universe (perhaps the universe we live). In section three we provide the necessary mathematical machinery, while in section four the energy spectrum is given. Section five is devoted to a physical application and in section six conclusions and final remarks are given.

\section{The model}

In this paper, we attempt to depict the effects of a curved spacetime on the spectrum energy of a test particle. In this way we are able to explore the behaviour of the aformentioned particle near the big bang, where quantum gravity effects are expected to be dominant.\\
Our metric background is provided by the Friedmann flat metric expressed in the usual comoving spatial coordinates
$\{x,y,z\}$ and the cosmic time $t$:
\begin{equation}
ds^2=-dt^2+a^2(t)\left(dx^2+dy^2+dz^2\right),
\label{1}
\end{equation}
with $a(t)$ the scale factor with the Hubble flow 
$H$ given by $H=\frac{da(t)}{dt}$. 
The so called concordance model is provided by a spatially flat metric equipped with a non-venishing
cosmological constant $\Lambda$ representing the $68\%$ of the actual energy content of our universe.\\
A widely used method to investigate (see for example \cite{2} and reference therein for a recent paper) the quantum nature of our 
primordial universe is provided by Wheeler-DeWitt equation. However,
shortcomings arise with the Wheeler-DeWitt equation
related in particular to the probabilistic interpretation of the wave function of the universe as a whole, since we have only one copy of our universe, and quantum mechanics requires an observer separated from the quantum system that can have at its disposal
a sets of copies of the quantum system. In this paper we use a more simple approach, in line with the idea of reproducibility of measures by an experimenter in a quantum context. In particular, we consider a test particle of mass $m$ in the metric (\ref{1}). Quantum effects are depicted in term of an external potential $V_{ext}$, where $V_{ext}$ represents quantum fluctuations of the universe as 'seen' by a given test particle with rest
mass $m$. As an example,
our universe could have zero total internal energy, with positive and negative energies
perfectly balancing at the apparent horizon. Moreover, in \cite{5} it has been shown, by using a generalization of the Bekenstein-Hawking entropy
\cite{6} that a zero internal energy at the apparent horizon results \cite{7} for our Friedmann flat universe. This open the possibility that our universe is born from 'nothing'. In such a scenario, we may suppose that test particle 'feels' a negative potential $V_{ext}<0$ with a total energy
$E=0$ in a universe with Minkowskian metric $a(t)=1$ and as a result a de Sitter universe with 
$a(t)=e^{c t\sqrt{\frac{\Lambda_i}{3}}}$ emerges, where $\Lambda_i$ represents the initial value of a primordial cosmological constant.\\ 
Another intriguing possibility is that non commutative effects at Planckian
scales modify the classical trajectory of $a(t)$ as dictated by
general relativity in such a way that the primordial big
bang singularity is ruled out \cite{I13}. This offers the possibility to test the allowed energy spectrum in a situation where gravity is strong enough to justify a quantum picture of the spacetime compared with a classical scenario where big bang is not ruled out; this will be studied in section five.
In the next two sections we show how the two cases, $E=0$ and $E<0$, can be obtained.\\
To start with, in the background (\ref{1}) we write the Schr\"{o}dinger equation using the Laplace-Beltrami Laplacian
${\nabla}^2_{\bf{g}}$. By denoting with $g$ the determinant of the spatial metric, obtained at $t=constant$, 
$g_{ij}$ the spatial metric ($\{i,j\}=\{1,2,3\}$) we have 
${\nabla}^2_{\bf{g}}=g^{-\frac{1}{2}}{\partial}_i\left(g^{ij}\sqrt{g}\;{\partial}_j\right)$. 
The Schr\"{o}dinger equation for the wave function
$\Psi=\Psi(t,x^i)$ of a quantum particle with rest mass $m$ interacting with an external potential $V_{ext}=V(t,x^i)$
becomes:
\begin{equation}
\imath\hbar\frac{\partial\Psi}{\partial t}=-\frac{{\hbar}^2}{2m}{\nabla}^2_{\bf{g}}\Psi+V_{ext}\Psi,
\label{2}
\end{equation}
with the Hamiltonian $\overline{H}$ given by $\overline{H}=\frac{p^2}{2m}+V_{ext}$. 
\\
First of all, we consider the free case with $V_{ext}=0$. The Beltrami-Laplace
operator is:
\begin{equation}
{\nabla}^2_{\bf{g}}=\frac{1}{a^2(t)}\left(
\frac{\partial^2}{\partial x^2}+\frac{\partial^2}{\partial y^2}+\frac{\partial^2}{\partial z^2}\right).
\label{3}
\end{equation}
The solution for (\ref{2}) with (\ref{3}) is given by:
\begin{equation}
\Psi(t, x^i)=\frac{1}{{(2\pi\hbar)}^{\frac{3}{2}}}
exp\left[\frac{\imath}{\hbar}\left(p_i x^i-\int\frac{p^2}{2ma^2(t)} dt\right)\right],
\label{5}
\end{equation}
where $p_i$ is the momentum with respect to the comoving coordinates $x^i$.
We must now specify a suitable expression for $V_{ext}$. Consider first a physical situation with $a(t)=1$, i.e. in a Minkowskian spacetime.
We may think to a test particle subjected to an attractive potential modeled by one attractive delta interaction located in a certain point, that for a sake of simplicity, we may think as the origin of the Cartesian coordinates:
$V_{ext}=\lambda \delta^3(\bf{x})$ with $\lambda<0$. As a consequence, our test particle is in a bound state if $E<0$ and a marginally bound state for $E=0$. In both cases, we may suppose that Planckian
fluctuations can produce an expanding universe. The suitable mathematical treatment of such a physical situations is provided in the next three sections. In a cosmological context, we may think to a 
$V_{ext}(t,x^i)$ looking like:
\begin{equation}
V_{ext}(t, x^i)=\frac{\lambda}{a^2(t)}\delta^3(\bf{x})=\overline{\lambda}\delta^3(\bf{x}).
\label{6}	
\end{equation} 
At the big bang, $t\rightarrow 0$ and $\overline{\lambda}\rightarrow -\infty$. For a universe without big bang, 
$\overline{\lambda}$ is finite $\forall t\in[0,+\infty)$. In any case, for late times $a(t)>>1$, where the delta interaction becomes negligible, a non primordial universe is expected with the ordinary classical description provided by the general relativity.
The chosen expression (\ref{6}) allows to study, thanks to the Laplace-Beltrami operator (\ref{3}), the energy configurations $E$ in the 
background (\ref{1}), setting $a(t)=1$ and performing the transformation $E\rightarrow E/a^2(t)$. In the next section we will adapt the treatment
in \cite{1} for the background (\ref{1}).

\section{Mathematical tools}
In this section we breafly exploit the mathematical machinery presented in \cite{1}, in order to depict quantum primordial fluctuations in terms of a delta potential interaction located, without loss of generality, at the origin of the Cartesian coordinates. The aformentioned interaction modelizes quantum effects that could be experienced by a test particle in a vacuum spacetime before the big bang. 
\\
As outlined in \cite{1} to get a rigorous mathematical treatment of a single delta point interaction one must check that the limit of two attractive delta interactions simmetrically situated, for example, at the zeta axis
$\pm \vec{z}_{0} =\left(\pm z_{0} ,0,0\right),z_{0} >0$, for $z_{0}\rightarrow 0$ is equivalent to a single point interaction with $\lambda\rightarrow 2\lambda$. The first step is to introduce a ultraviolet cutoff $k>0$: $\left|\vec{p}\right|=\sqrt{p_{x}^{2} +p_{y}^{2} +p_{z}^{2} } \le k,$ \footnote{In this section for sake of semplicity we use geometrized unit with $\hbar=m=1$. The usual units will be restored in the next section.}.\\
Concerning the resolvent limit, as $k\rightarrow +\infty$
describing the Hamiltonian (\ref{3}) with $a(t)=1$ and
with cut-off  \textit{k} and coupling constant $\lambda(k)\neq0$ is:
\begin{eqnarray}
& &	H(k,z_{0} )=\left|\vec{p}\right|^{2} +\label{aa}\\
& &+\frac{\lambda (k)}{(2\pi )^{3} } \left[|\chi _{\left|\vec{p}\right|\le k} e^{-i\vec{z}_{0} \cdot \vec{p}} ><\chi _{\left|\vec{p}\right|\le k} e^{-i\vec{z}_{0} \cdot \vec{p}} |+|\chi _{\left|\vec{p}\right|\le k} e^{i\vec{z}_{0} \cdot \vec{p}} ><\chi _{\left|\vec{p}\right|\le k} e^{i\vec{z}_{0} \cdot \vec{p}} |\right]\nonumber
\end{eqnarray}
The resolvent for $E<0$ by means \cite{P1} is given by:
\[\left[H(k,z_{0} )+|E|\right]^{-1} =\left[\left|\vec{p}\right|^{2} +|E|\right]^{-1}-\]
\[\frac{\frac{2}{(2\pi )^{3} } |\chi _{\left|\vec{p}\right|\le k} \cos (\vec{z}_{0} \cdot \vec{p})\cdot (\left|\vec{p}\right|^{2} +|E|)^{-1} ><\chi _{\left|\vec{p}\right|\le k} \cos (\vec{z}_{0} \cdot \vec{p})\cdot (\left|\vec{p}\right|^{2} +|E|)^{-1} |}{\frac{1}{\lambda (k)} +\frac{2}{(2\pi )^{3} } ||\chi _{\left|\vec{p}\right|\le k} \cos (\vec{z}_{0} \cdot \vec{p})\cdot (\left|\vec{p}\right|^{2} +|E|)^{-1/2} ||_{2}^{2} } -\] 

\begin{equation} 
	\frac{\frac{2}{(2\pi )^{3} } |\chi _{\left|\vec{p}\right|\le k} \sin (\vec{z}_{0} \cdot \vec{p})\cdot (\left|\vec{p}\right|^{2} +|E|)^{-1} ><\chi _{\left|\vec{p}\right|\le k} \sin (\vec{z}_{0} \cdot \vec{p})\cdot (\left|\vec{p}\right|^{2} +|E|)^{-1} |}{\frac{1}{\lambda (k)} +\frac{2}{(2\pi )^{3} } ||\chi _{\left|\vec{p}\right|\le k} \sin (\vec{z}_{0} \cdot \vec{p})\cdot (\left|\vec{p}\right|^{2} +|E|)^{-1/2} ||_{2}^{2} }
\end{equation} 
It's crucial to notice that, a necessary and sufficient condition, in order to obtain a finite expression after removing the cut-off $k$ ($k\rightarrow +\infty$), is provided by the expression: 
\begin{equation}
\lambda (k,\beta )=-\frac{\beta }{1+\frac{1}{(2\pi )^{3} } \int _{p|\le k}^{}\frac{1}{\left|\vec{p}\right|^{2} } d^{3} p }.
\label{rc}
\end{equation} 
It's important to point out that in formula (\ref{rc}) $\lambda$ represents the bare non interacting coupling constant, while $\beta$ denotes the interacting one appearing in the renormalized Hamiltonian.
As shown in \cite{P1}, {\it Theorem 2.1}, with this procedure a self-adjoint Hamiltonian is obtained.
For the ground state and the first excited bound state energy we have the following expressions:
\begin{eqnarray}
& & \alpha +\frac{|E|^{1/2} }{4\pi } -\frac{1}{4\pi } \frac{e^{-2|E|^{1/2} z_{0} } }{2z_{0} } =0,\alpha =\frac{1}{\beta } ,
\label{gs}
\\
& & \alpha +\frac{|E|^{1/2} }{4\pi } +\frac{1}{4\pi } \frac{e^{-2|E|^{1/2} z_{0} } }{2z_{0} } =0.
\label{bs}
\end{eqnarray}
By taking the limit $z_{0}\rightarrow 0$, the magnitude of the ground state energy diverges. In order to correct this singular behavior we need to regularize the expression of the coupling constant, in such a way that this must be dependent also by the choice of $z_{0}$:
\begin{equation}
\lambda (\beta ,z_{0} )=-\frac{\beta }{1+\frac{1}{(2\pi )^{3} } \int _{p}^{}\frac{1+\cos (2\vec{z}_{0} \cdot \vec{p})}{\left|\vec{p}\right|^{2} } d^{3} p }.
\end{equation}
Finally we are able to write the renormalized Hamiltonian with the interacting coupling constant $\beta$
\begin{equation}
H_{\left\{\beta ,\vec{z}_{0} \right\}} =-\Delta +\beta\left[\delta (\vec{z}-\vec{z}_{0} )+\delta (\vec{z}+\vec{z}_{0} )\right].
\label{Ham}
\end{equation}
We are now in the position to write down and study, in the next section, the explicit espression for the ground state energy and first excited one.

\section{Spectrum analysis}

As a consequence of setups of the above section, after restoring the usual unit measure for the energy of the lower bound, we get:
\begin{equation}
\alpha +\frac{\sqrt{2m|E|}}{4\pi\hbar} +\frac{1}{4\pi}\frac{\left(1-e^{-\frac{2 z_{0}\sqrt{2m|E|}}{\hbar}}\right)}{2z_{0}} =0,\; \alpha =\frac{1}{\beta}, 
\label{GS}
\end{equation}
while for the first excited bound state:
\begin{equation}
\alpha +\frac{\sqrt{2m|E|}}{4\pi\hbar} +\frac{1}{4\pi}\frac{\left(1+e^{-\frac{2 z_{0}\sqrt{2m|E|}}{\hbar}}\right)}{2z_{0}} =0,\; \alpha =\frac{1}{\beta}, 
\label{st}
\end{equation}
The above equations (\ref{GS}) and (\ref{st}) are thus the regularized formulas respectively for the ground state and the bound state
providing the suitable limit $z_0\rightarrow 0^+$. Note that 
the ground state energy gets absorbed into the absolutely continuous spectrum exactly at \textit{$\alpha $}=0, 
with $E_{0}(\alpha=0)=0$. This is in agreement with the fact that this operator should approach, as \textit{z}$ _{0}\rightarrow 0_{+} $, the negative Laplacian perturbed by a single point interaction centred at the origin which is known to have a \textit{zero energy resonance} at \textit{$\alpha $}=0.
\\
For the norm resolvent limit of H(\textit{$\beta $,z}$ _{0} $), as \textit{z}$_{0}\rightarrow 0_{+}$, we obtain:
\begin{equation}
	\left[\left|\vec{p}\right|^{2} +|2mE|/\hbar^2\right]^{-1} +\frac{\frac{1}{(2\pi )^{3} } |\frac{1}{\left|\vec{p}\right|^{2} +|2mE|/\hbar^2} ><\frac{1}{\left|\vec{p}\right|^{2} +|2mE|/\hbar^2} |}{\frac{1}{2\beta } +\frac{|2mE|^{1/2} }{4\pi\hbar } } \label{res}
\end{equation} 
Formula (\ref{res}) represents the resolvent \cite{P1} in momentum space of the negative Laplacian perturbed by a single point interaction centred at the origin having double strength. 
\\
It is worth to be noticed that for $\beta <0$ is also present an isolated eigenvalue, below the absolutely continuous spectrum, given by
\begin{equation}
E_{0} (2\beta )=-\left(\frac{2\pi }{\beta} \right)^{2}\frac{\hbar^2}{2m}.
\label{br}
\end{equation}
As clarified at the end of section two, in order to obtain the expression (\ref{br}), suitable for the background metric (\ref{1}), the transformation $E\rightarrow E/a^2(t)$ is necessary; we thus obtain
\begin{equation}
E_{0}(2\beta,t)=-\left(\frac{2\pi }{\beta} \right)^{2}\frac{\hbar^2}{2m\;a^2(t)}.
\label{br1}
\end{equation}
Thanks to the formula (\ref{br1}), we can test the effects of an attractive $\delta$-interaction in an expanding Friedmann flat universe on a given test particle, thus extending the treatment in \cite{P1} in a spacetime provided with a non-vanishing curvature.

\section{An application: testing a primordial Friedmann universe on a non commutative spacetime}

Due to all preliminary observations above, note that for a Friedmann solution at late times, where 
for $t\rightarrow +\infty$ we have $a(t)\rightarrow +\infty$, so we get $E_{0} (2\beta )\rightarrow 0$: this implies that, thanks to the effects of the expanding universe, the energy of the test particle belongs to the absolute continuous spectrum obtained in \cite{P1} for $\beta\rightarrow +\infty$. Moreover, for $\beta\rightarrow +\infty$ in (\ref{br1}), $E_{0} (2\beta )\rightarrow 0$ independently of the value of $a(t)$, also for big bang cosmology, where $a(t)\rightarrow 0$ for $t\rightarrow 0$. These facts can be of interest in light of the results in
\cite{5}, where the possibility that the universe is born from 'nothing', i.e. from s state with $E=0$, is explorated. 
\\
The case where the ground state energy is strictly negative is much more intriguing. In fact, if $a(t)$ follows the classical trajectory at the big bang singularity, then $E_{0} (2\beta )\rightarrow -\infty$; this means that the ground state energy is not bounded from below, as a result the system becomes unstable and finally collapses. This picture changes drastically if $a(t)$ does not follow the classical trajectory, this is possible  when quantum gravity planckian fluctuations come in action at Planckian scales. This can be 
physically possible by supposing a quantum spacetime at Planckian scale \cite{I8,I9,I10,I11,I12,I13}. In particular in \cite{I8} a quantum spacetime model has been obtained starting from physically motivated uncertainity relations (STUR), finding commutation relations among spacetime coordinates 
$q^{\mu}$. Moreover, the commutation relations in \cite{I8} generate a non commutative
$C^{*}$ algebra $\mathcal{\epsilon}$
acting on a generic Hilbert space $\mathcal{H}$. In this way the spacetime coordinates $\{q^{\mu}\}$ are elevated to selfadjoint operators, where the classical Poicaré simmetry is implemented at a quantum level to the commutation rules $[q^{\mu},q^{\nu}]=\imath\hbar Q^{\mu\nu}$, with $Q^{\mu\nu}$ a covariant tensor 
only under proper proper Lorentz transformations. 
The construction of a realistic non-commutative model in a curved spacetime, where the diffeomorphism group transformation among coordinates acts at classical level, represents a formidable task. As noticed in \cite{I9, I11}, in order to accomplish this issue, the following system must be solved:
\begin{eqnarray}
& & [q^{\mu},q^{\nu}]=\imath\hbar Q^{\mu\nu}(g_{\mu\nu}), \label{s1}\\
& & R_{\mu\nu}-\frac{1}{2}R g_{\mu\nu}=T_{\mu\nu}(\psi),\label{s2}\\
& & F(\psi)=0, \nonumber
\end{eqnarray}  
where the energy momentum tensor is supposed to depend on some fields $\psi$ with their equation of motion.
Since the commutators in (\ref{s1}) depend on the background metric $\bf g$ that in turn depends on the coordinates
$\{q^{\mu}\}$, with the Einstein equations depending on $\bf g$, the system (\ref{s1}) is quite difficult to solve.
As stated in \cite{I9}, a more easy way to attack the (\ref{s1}) is to consider a semi-classical approximation where instead of 
$T_{\mu\nu}$ we have its expectation value on some allowed state $\{\omega\}$ i.e. ${<T_{\mu\nu}>}_{\omega}=\omega(T_{\mu\nu})$ and with
$\bf g$ the classical solution. To this purpose, in \cite{I13}, a non commutative spacetime in a Friedmann flat background has been obtained.
The starting point are the STUR that in \cite{I13} have been given in term of the proper spatial variables 
${\eta}^i= a(t) x^i$ and the time one $t$. We denote with 
$\Delta_s A=\sum_{i,j, i\leq j}\Delta_s {\eta}^i \Delta_s {\eta}^j$ and 
$\Delta_s V=\prod_i\Delta_s {\eta}^i$, where $s$ denotes a generic state. One get \cite{I13}:
\begin{eqnarray}
& & \frac{\sqrt{\Delta_s A}}{4\sqrt{3}}+\frac{s(H)\Delta_s A}{12c}\geq \frac{L_P^2}{2} 
\frac{\Delta_s A}{\Delta_s V}, \label{s3}\\
& & c\Delta_s t\left(\frac{\sqrt{\Delta_s A}}{4\sqrt{3}}+\frac{s(H)\Delta_s A}{12c}\right)\geq \frac{L_P^2}{2},\nonumber
\end{eqnarray}
where the dependence on the state $s$ is explicitely indicated.
In a quantum spacetime \cite{I8}, only pure states are considered that are maximally localizing, i.e. states saturing the STUR.
These states are the ones with spherical symmetry with 
$\Delta_s {\eta}^i\sim \Delta_s {\eta}^j\sim x\Delta_s t =\Delta_s {\eta}$, i.e. the uncertainties are all of the same magnitude.
In order to obtain the (\ref{s1}) implying the STUR, the following weaker version has been used in \cite{I13}:
\begin{eqnarray}
& & \Delta_s A\left(\frac{1}{4\sqrt{3}}+\frac{s(H)\sqrt{\Delta_s A}}{12 c}\right)\geq \frac{L_P^2}{2}, \label{st2}\\
& & c\Delta_s t\left(\Delta_s \eta_1 +\Delta_s \eta_2 +\Delta_s \eta_3\right)\left(\frac{1}{4\sqrt{3}} +\frac{s(H)\sqrt{\Delta_s A}}{12 c}\right)\geq \frac{L_P^2}{2}.\nonumber
\end{eqnarray}
In \cite{I3} a concrete realization of a quantum spacetime has been provided with the coordinates $\{t,{\eta}^i\}$
(and so also $\{t,x^i\}$) promoted to essentially self-adjoint operators $\{\hat{t},{\hat{x}}^i\}$
satisfying the STUR (\ref{st2}), together with
 a (strongly continuous) unitary representation of the global isometry group $G$ under which the operators $\eta$ transform as their classical counterparts. For the purposes of this paper, the exact form of a concrete realization of the commutators among
 the coordinates $\{t, x^i\}$ is not necessary: the important point is the existence of a concrete model with commutators in (\ref{s1})
 implying the STUR (\ref{s3}) or their weaker realization (\ref{st2}).\\
 The next step to attack the system (\ref{s1})-(\ref{s2}) is to obtain a semi-classical texture of Einstein equations. In a Friedmann background
 with the classical metric (\ref{1}), the classical Einstein equations will depend on the matter energy content of the spacetime, summarized by
 $T_{\mu\nu}$, and on the metric factor $a(t)$. In a semi-classical treatment, $T_{\mu\nu}$ must be substituted in Einstein equations
 with its mean value $s(T_{\mu\nu})$, with a(t) a function fulfilling the semi-classical Einstein equations with $t$ and $x^i$
 in (\ref{1}) belonging respectively to the spectrum of $\hat{t}$ and ${\hat{x}}^i$, i.e. 
 $t\in\;sp(\hat{t})$ and $x^i\in\;sp({\hat{x}}^i)$. Also a semi-classical texture of (\ref{s2}) is not a simple task. In \cite{6} we have studied a possible effective way to depict the modifications on the Einstein equations caused by Planckian fluctuations. In \cite{6} it has been shown that Planckian effects produce fluctuations $U$ on the energy content of the universe looking like
 $U\sim c\hbar\overline{c_0}\sqrt{\Gamma}$, where $\overline{c_0}\in\mathcal{R}$, and $\Gamma$ has the dimensions of the inverse length and it is a measure of the areal radius of the spherical region under consideration: for example for a universe equipped with a non-vaniahing cosmological constant $\Lambda$ we have $\Lambda\sim\Gamma$, with $1/\sqrt{\Gamma}$ of the order of the Hubble radius (apparent horizon) 
 in a classical de Sitter spacetime. When these fluctuations are taken into account, a Planckian texture for a semi-classical Friedmann flat universe can be obtained \cite{6}. Finally, the so motivated version of the system (\ref{s1})-(\ref{s2}) in our semi-classical approximation depicting a quantum spacetime with a test particle is:
\begin{eqnarray}
& & [\hat{t},\hat{\eta}^j ]=2i\hbar f\circ\left(H(\hat{t})\right),\;\;[\hat{\eta}^i, \hat{\eta}^j ]= 2i\hbar f\circ H(\tilde{t}),\;
i\neq j,\label{s4}\\
& &{\hat{\eta}}^i=\frac{a(\hat{t}){\hat{x}}^i+{\hat{x}}^ia(\hat{t})}{2},\label{s5}\\		
& & f^3 + k f^2 - 2\sqrt[4]{3}k = 0,\;\;\;k= \sqrt{4c/3 L_P H(t)},\;H(t)\geq 0, \label{s6}\\
& & {H}^2(t)=\frac{c^2\sqrt{3}}{4{\overline{c_0}} L_P^2}
\left[-1+\sqrt{1+\frac{64\pi{\overline{c_0}}}{3\sqrt{3}}\frac{s(\rho)}{{\rho}_P}}\right],\;
{\rho}_P=\frac{c^5}{\hbar G^2},
\label{s7}\\
& & c^2\left(1+\frac{4\overline{c_0}L_P^2}{c^2\sqrt{3}}H^2\right)H_{,t}=-4\pi G(c^2{s(\rho)}+
{s(p)}), \label{s8}\\ 
& & \overline{H}=
{\nabla}^2_{\bf{g}}=\frac{1}{a^2(t)}\left(
\frac{\partial^2}{\partial x^2}+\frac{\partial^2}{\partial y^2}+\frac{\partial^2}{\partial z^2}\right)
+V_{ext}(t, x^i).
\label{s9}
\end{eqnarray}
The system (\ref{s4})-(\ref{s9}) represents our semi-classical texture of the system (\ref{s1})-(\ref{s2}) and depicts a test particle
in a non commutative spacetime. The (\ref{s4}) represent the commutation relations among the coordinate operators 
$\{\hat{t}, {\hat{\eta}}^i\}$, that can be obtained \cite{I13} by means of the covariantisation trick \cite{8}. Commutators are valid
for any state $s$ and do imply the  STUR (\ref{st2}). Formula
(\ref{s5}) is the usual prescription to extend the classical expression ${\eta}^i=a(t) x^i$ to a quantum non commutative level, while 
the (\ref{s6}) denotes the algebraic equation satisfied by $f$ in (\ref{s4}). The (\ref{s7})-(\ref{s8}) are the Friedmann flat equations
with Planckian fluctuation corrections\footnote{Similar equations have been obtained in \cite{7} in 
a holographic braneworld scenario}. It is easy to see that equations  (\ref{s7})-(\ref{s8}) reduce to the classical ones of general relativity
sufficiently away from the classical big bang singularity where $\frac{s(\rho)}{{\rho}_P}<<1$ and $H(t)<<1/t_P$, with
$t_P$ the Planck time. This is according to the fact that non commutativity is very strong near the classical spacelike singularity.\\
Finally, equation (\ref{s9}) is the Hamiltonian of our test particle with the potential 
$V_{ext}$ given by (\ref{6}). Equations (\ref{s7})-(\ref{s9}) are defined in terms of the comoving coordinates
$\{t, x^i\}$, where $t\in sp(\hat{t})$, $x^i\in sp({\hat{x}}^i)$ with $t\in\mathcal{R}^+$, $x^i\in\mathcal{R}$. For 
(\ref{s7})-(\ref{s8}), the energy density $\rho$ and the pressure $p$ enter in the equations as mean value in a state $s$. Note that
in (\ref{s7})-(\ref{s8}) the sign of the constant $\overline{c_0}$ has left unspecified. To this purpose, an expected feature of Friedmann equations capturing non commutative effects is that the initial spacelike big bang singularity at $t=0$ is ruled out. It is a simple matter to verify that only with $\overline{c_0}<0$ big bang is avoided. In fact, for $\overline{c_0}<0$ equation (\ref{s7}) implies a maximum allowed value
for $s(\rho)$, and consequently for $H(t)$, given by:
\begin{equation}
{s({\rho})}\leq {s({\rho})}_{max}=\frac{3\sqrt{3}{\rho}_P}{64\pi|\overline{c_0}|},\;\;
H\leq {{H}}_{max}=\frac{c\;3^{\frac{1}{4}}}{2\sqrt{|\overline{c_0}|}\;L_P}.
\label{s10}
\end{equation}
In the Einstein equations, $a(t)\rightarrow 0$ for $t\rightarrow 0$ with 
$H(t\rightarrow 0)=\rho(t\rightarrow 0)=+\infty$. It is interesting to note that the inequalities (\ref{s10}) hold for any physical state
$s$ and is a consequence of the form of (\ref{s7})- (\ref{s8}). Concerning the value of  $\overline{c_0}$, it is not fixed in our model. 
However, since we expect that $ {s({\rho})}_{max}\leq {\rho}_P$ we expect 
$|\overline{c_0}|\geq\frac{3\sqrt{3}}{64\pi}$. As an example, in loop cosmology \cite{9}, we have 
${s({\rho})}_{max}\simeq 0.41 {\rho}_P$, obtained in our case for $|\overline{c_0}|\simeq 0.06$
\footnote{Note that in our model the big bang singularity is not avoided with a bounce behavior, as happens in loop cosmology.}.\\
As an istructive example, we analyze the case with $s({\rho})=\frac{3k}{8\pi G t^2}$: this corresponds to the classical solution with
a power law cosmology with $a(t)\sim t^{k}$ and $k\in(0, +\infty)$. Obviously, equation (\ref{s7}) implies that also with 
$s(\rho)$ mimicking the classical trajectory, the solution for $a(t)$ approaches the classical solution sufficiently away from the strong gravitational regime near the Planck time $t_P$, but it is affected by non commutative effects near the Planck time.
In fact, after putting $s({\rho})=\frac{3 k}{8\pi G t^2}$ in (\ref{s7}) we obtain:
\begin{equation}
{H}^2(t)=\frac{c^2\sqrt{3}}{4{|\overline{c_0}|} L_P^2}
\left[1-\sqrt{1-\frac{8 k|\overline{c_0}|}{\sqrt{3}}\frac{L_P^2}{c^2 t^2}}\right].
\label{s11}
\end{equation}
From a first inspection of (\ref{s11}), note that the solution cannot be extended up to the classical singularity at $t=0$. In
fact, a minimum time $t_{min}$ arises given by:
\begin{equation}
t_{min}=t_P\sqrt{\frac{8}{\sqrt{3}} k|\overline{c_0}|}.
\label{s12}                                                      
\end{equation}                                                   
For a classical universe filled with radiation, i.e. $k=\frac{1}{2}$, we have that $t_m\sim t_P$
for $|\overline{c_0}|\sim 0.4$. This means that the solution of (\ref{s11}) is defined for 
$t\in[t_{min}, +\infty)$. The formal solution of (\ref{s11}) is:
\begin{eqnarray}
& & a(t)=a_{min} e^{\beta(t)}, \label{s13}\\
& & \beta(t)=\sqrt{\frac{c^2\sqrt{3}}{4{|\overline{c_0}|} L_P^2}}\int_{t_{min}}^{t}
\sqrt{\left[1-\sqrt{1-\frac{8 k|\overline{c_0}|}{\sqrt{3}}\frac{L_P^2}{c^2{t^{\prime}}^2}}\right]}dt^{\prime}.\nonumber
\end{eqnarray}
For $t>>t_{min}$ in (\ref{s13}) the classical power law behavior $a(t)\sim t^{k}$ is regained.
Note that for $|\overline{c_0}|\sim 0.01$ or greater but less than $a$, with $a$ of the order of unity, we have that 
$t_{min}\sim t_P$ with obviously $a_{min}\in\mathcal{R^+}$ and as a consequence the classical big bang singularity 
at $t=0$ is ruled out, also by taking an expression for $s({\rho})$ according to the classical trajectory. This picture is in perfect agreement with the one expected for a quantum spacetime in a strong regime. In fact, the STUR (\ref{st2}), that are implied by the commutation relations
(\ref{s4}), for maximally localized spherical states with $H_{max}$ given by (\ref{s10}), do imply that 
$\Delta_s t\sim t_P$. Thus an expanding universe emerges at $t\sim t_P$.\\
To complete our semi-classical picture, we must discuss the last equation (\ref{s9}), i.e. of our test particle. A complete (quantum field theory QFT) description of a quantum spacetime requires the introduction of quantum fields, as in \cite{I8}. For the presence of time dependence in the background metric, this is a yet unsolved task (see for example \cite{I9,I11} and also \cite{I13}). Thanks to
the introduction of quantum field  $\Phi(S)$, we can obtain
all quantum observables localised in a region $\cal O$, with $S$ a Schwartz test function. In particular, particles can be rigorously defined
in terms of quantum field acting on some Hilbert space $\mathcal{H}$. A semi-classical way to treat test particles in a quantum spacetime is provided by the system (\ref{s4})-(\ref{s9}). The universe at the Planck time can be seen as a 'vacuum' where quantum Planckian fluctuations come into action and a Friedmann universe emerges without the classical singularity. As stated above equation (\ref{s4}), Planckian fluctuations
with $\overline{c_0}<0$ leads to a negative energy $U<0$. Such a negative energy, experienced by our test particle, is represented in our model by an actractive $3D$ delta interaction. We may think for example 
to a 'foam' of particles (perhaps bosons) created in the vacuum state interacting with our test particle with an attractive potential (negative fluctuations with $\overline{c_0}<0$). Mathematically, the quantum spacetime is represented by 
(\ref{s4}), with the semi-classical Friedmann equations given by (\ref{s7})-(\ref{s8}). Test particle is equipped with its wave function
$\Psi(t, x^i)$ and Hamiltonian given by the (\ref{s9}) that in turn is coupled to the quantum spacetime by the scale factor $a(t)$, satisfying equation
(\ref{s11}) with $t\in sp(\hat{t})\in [t_{min}, +\infty)$, $x^i\in sp({\hat{x}}^i)\in\mathcal{R}$. Therefore, in the primordial universe our test particle, thanks to the Planckian fluctuations with negative energy, is in its ground state with energy given by:
\begin{eqnarray}
& & E_{0}(2\beta,t)=\label{s14}\\
& & -\left(\frac{2\pi }{\beta} \right)^{2}\frac{\hbar^2}{2m\;a^2(t)}\geq
-\left(\frac{2\pi }{\beta} \right)^{2}\frac{\hbar^2}{2m\;a_{min}^2(t)},\;\;t\in[t_{min}, +\infty).
\nonumber
\end{eqnarray}
As a physical consequence of (\ref{s14}), the presence of the ground state energy for test particle is an outcome of the quantum nature of the spacetime at Planckian scales. In a spacetime where $a(t)$ follows the classical trajectory up to the big bang singularity at 
$t=0$, such a ground state cannot exist. At late times with $t>>t_{min}$ the classical situation is regained with 
$E_{0}(2\beta,t\rightarrow +\infty)\rightarrow 0$ and the ground state of the test particle is reabsorbed in the absolutely
continuum spectrum as happens
in \cite{P1} for $\beta\rightarrow +\infty$.

\section{Conclusions and final remarks}

In this paper, we have proposed a model to describe possible quantum gravity effects near the classical big bang singularity.
We used the resuts in \cite{P1} in order to obtain the spectrum of a test particle in a Friedmann flat spacetime
(perhaps the universe we live). As a first result, with $a(t)$ fulfilling the classical trajectory with the big bang singularity 
at $t=0$, the ground state with negative energy is not allowed. The situation drastically changes if Planckian fluctuations of the geometry are taken into account, this is the case when a non commutative geometry is introduced in the classical Friedmann flat background. To obtain a complete quantum spacetime, the system (\ref{s1}), as proposed in \cite{I9,I11}, must be solved. Unfortunately, nobody has been yet able to solve this complicated system. To this purpose we propose, as suggested in \cite{I9}, a semi-classical texture of (\ref{s1}), namely
system (\ref{s4})-(\ref{s9}). There, the background metric is depicted in terms of continuous functions of the
coordinates $\{t, x^i\}$, with  $t\in sp(\hat{t})$, $x^i\in sp({\hat{x}}^i)$ with $t\in\mathcal{R}^+$, $x^i\in\mathcal{R}$.
The operators $\hat{t}$ and ${\hat{x}}^i$ satisfy the commutation relations (\ref{s4}), that in turn imply the STUR
(\ref{st2}). As an outcome of this semi-classical picture, the initial big bang singularity is ruled out and a maximum energy density together with Hubble flow emerges (\ref{s10}). As a consequence, our test particle, since fluctuations produce negative energies, experiences a
ground state with finite negative energy modeled by the Hamiltonian (\ref{s9}). Hence, such a primordial ground state energy can be created near 
$t=t_P$ only thanks to the non commutative spacetime nature.\\
Apart from the interesting phenomenon depicted above, the system (\ref{s4})-(\ref{s9}) can represent an usefull concrete way to obtain a semi-classical description of (\ref{s1})-(\ref{s2}). 
In particular, the (\ref{s7})-(\ref{s8}) represents semi-classical Friedmann flat equations with Planckian fluctuations within.\\
As well known, in order to depict particles on a curved quantum spacetime, a QFT, realized in \cite{I9} for a Minkowskian 
background, is necessary. Ufortunately this QFT is not, at present day, at our disposal. To overcome this issue, we depict Planckian fluctuations
experienced by a test particle in terms of the Hamiltonian (\ref{s9}), coupled to a quantum spacetime by the scale factor $a(t)$.
In this way, the ground state with energy (\ref{s14}) emerges. In a Friedmann context, the ground state acquires a time dependence and it is monotonically increasing, approaching zero from negative values.\\
Awaiting for a generally accepted quantum gravity proposal, this paper is an attempt to find possible physical effects arising when quantum systems are embedded in strong gravitational fields, in particular in a cosmological context, by using sound arguments of quantum mechanics, 
non commutative geometry, operator theory and general relativity, summarized by the system (\ref{s4})-(\ref{s9}), that in 
turn is a concrete semi-classical realization of the system (\ref{s1}) proposed in \cite{I9,I11}. A possible future line
of research could be to use particular external potentials $V_{ext}$ or more general Hamitonian than (\ref{s9}) in the system
(\ref{s4})-(\ref{s8}).

\end{document}